\documentclass{aastex62}
\usepackage{bm}
\usepackage{amsmath}
\usepackage{mathrsfs}

\received{}
\revised{}
\accepted{}

\submitjournal{ApJ}

\shorttitle{Gamma-rays and Neutrinos from SNe II}
\shortauthors{Wang et al.}

\begin{document}

\title{Transient High-energy Gamma-rays and Neutrinos from Nearby Type II Supernovae}

\correspondingauthor{Kai Wang}
\email{kaiwang@pku.edu.cn}

\author{Kai Wang}
\affiliation{Department of Astronomy, School of Physics, Peking University, Beijing 100871, China}
\affiliation{Kavli Institute for Astronomy and Astrophysics, Peking University, Beijing 100871, China}

\author{Tian-Qi Huang}
\affiliation{Department of Astronomy, School of Physics, Peking University, Beijing 100871, China}
\affiliation{Kavli Institute for Astronomy and Astrophysics, Peking University, Beijing 100871, China}

\author{Zhuo Li}
\affiliation{Department of Astronomy, School of Physics, Peking University, Beijing 100871, China}
\affiliation{Kavli Institute for Astronomy and Astrophysics, Peking University, Beijing 100871, China}

\begin{abstract}
The dense wind environment (or circumstellar medium) may be ubiquitous for the regular Type II supernovae (SNe) before the explosion, the interaction of which with the SN ejecta could result in a wind breakout event. The shock generated by the interaction of the SN ejecta and the wind can accelerate the protons and subsequently the high-energy gamma-rays and neutrinos could arise from the inelastic $pp$ collisions. In this work, we present the detailed calculations of gamma-ray and neutrino production for the regular Type II SNe. The calculation is executed by applying time-dependent evolutions of dynamic and proton distribution so that the emission could be shown at different times. Our results show, for the SN 2013fs-like wind environment, the multi-GeV and $\sim \mathrm{few}-100 \,\rm TeV$ gamma-rays are detectable with a time window of several days at $\lesssim 2-3\,\rm Mpc$ by \emph{Fermi}/LAT and CTA during the ejecta-wind interaction, respectively, and can be detected at a further distance if the wind environment is denser. Besides, we found the contribution of the wind breakouts of regular Type II SNe to diffuse neutrino flux is subdominant by assuming all Type II SNe are SN 2013fs-like, whereas for a denser wind environment the contribution could be conspicuous above $300\,\rm TeV$.
\end{abstract}


\section{Introduction}\label{s1}
Type II supernovae (SNe) originate from the explosion of hydrogen-rich supergiant massive star. For Type IIn and superluminous SNe, the massive stars may experience mass-lose episodes before they explode as SNe to form a dense wind environment (or circumstellar medium) \citep{smith07, miller09, ofek13a, ofek13b, margutti14}. Recently, thanks to the rapid follow-up spectroscopy observations, SN 2013fs, a regular Type II SN, is suggested to be with a dense wind environment that is produced by the progenitor prior to explosion at a high mass-loss rate $\sim 3 \times 10^{-3} (v_{w}/100\,\mathrm{km\,s^{-1}}) \,\rm {M_ \odot }\,{yr^{ - 1}}$, where $v_w$ is the assumed velocity of wind \citep{yaron17}. Moreover, very recently, the further early observations for dozens of rising optical light curves of SNe II candidates indicate that the SN 2013fs is not a special case and the circumstellar wind materials should be ubiquitous for regular Type II SNe \citep{forster18}. The pre-explosion mass-loss can be with a rate of $\dot M > 10^{-3}(v_{w}/100\,\mathrm{km\,s^{-1}})\,\rm {M_ \odot }\,{yr^{ - 1}}$ and last for years. So universally, for the regular Type II SNe, the circumstellar wind environment with a high density in the immediate vicinity of the progenitor can be caused by the sustained mass-loss of the progenitor before the explosion.

After the SN II explosion, the interaction of SN ejecta with the optically thick wind could result in a bright, long-lived wind breakout event, which may also make the usual envelope breakout delay. The shock generated by the interaction of the SN ejecta and the wind can accelerate the protons, and subsequently the inelastic $pp$ collision between the accelerated protons and the shocked wind gas can give rise to the signatures of neutrinos and gamma-rays \citep{katz11, murase11, zirakashvili16, petropoulou17}. The produced neutrinos and gamma-rays could be the crucial probe to identify the problems of these explosive phenomena, e.g., the properties of progenitor and the acceleration of cosmic rays \citep{murase14}. In addition, the produced neutrinos could contribute to the diffuse neutrino emission \citep{li18}.

In this work, we have presented the gamma-ray and neutrino emissions during the ejecta-wind interaction by focusing on the regular Type II SNe that are with a much higher event rate than Type IIn and superluminous SNe, and the typical values of parameters based on the SN 2013fs are adopted. The dynamic of ejecta-wind interaction is calculated with a time-dependent (radius-dependent) evolution and the gamma-ray and neutrino emissions are presented through the detailed calculations. The modification of proton distribution due to the cooling and injection at different radii are taken into account. Besides, since the dense wind environment is common for the regular Type II SNe, we derive the diffuse neutrino flux contributed by the SNe II wind breakouts by assuming all Type II SNe are 2013fs-like. This paper is organized as follows. In Section \ref{s2}, we describe the dynamics of SN ejecta-wind interactions. We calculate the different timescales of protons in the shocked wind region in Section \ref{s3} and present the gamma-ray and neutrino emission in Section \ref{s4} as well as the contribution to diffuse neutrino emission. Discussions and conclusions are given in Section \ref{s5}.

\section{Dynamics}\label{s2}

The SN explosion ejects the progenitor's stellar envelope. Typically the ejecta is with a bulk kinetic energy of $E_k=10^{51}\mathscr{E} \,\rm{erg}$ and an total mass of $M_{ej}=10 \mathscr{M} {M_ \odot }$, inducing a bulk velocity ${v_b} = \sqrt {2{E_k}/{M_{ej}}}  = 3.2 \times {10^8}{\mathscr{E}^{1/2}}{\mathscr{M}^{ - 1/2}}\,\rm cm \,{s^{ - 1}}$. After shock breakout from stellar envelope, the energy of ejecta with velocity larger than $v$ can be described by \citep{matzner99, li18}
\begin{equation}
E( > v) = {E_k}{(v/{v_b})^{ - \chi }}
\end{equation}
where $\chi  = 3+5/n$. For the convective (radiative) envelopes, one has $n=3/2$ $(3)$ and $\chi= 19/3$ $(14/3)$ \citep{matzner99}. In this work, we adopt $\chi =6$ for RSGs (red supergiant stars), while we also try $\chi =5$ for BSGs (blue supergiant stars) which gives a negligible difference.

After shock breakout from the stellar envelope, the interaction of the SN ejecta with the wind forms a forward shock with velocity $v_s$ to propagate through the wind and a reverse shock to cross the SN ejecta. Since the gamma-ray and neutrino emission of reverse shock are usually weaker than that of the forward shock \citep{murase11}, we neglect the contribution of the reverse shock. For the regular Type II SN 2013fs-like case, according to the measurement the wind profile is suggested as $\rho(R)=A R^{-2}$ with $A = \dot M/4\pi {v_w} = 1.5 \times {10^{15}}\mathscr{A} \,\rm g \, cm^{ - 1}$ for $\dot M = 3 \times {10^{ - 3}}\,\rm {M_ \odot }\,{yr^{ - 1}}$ and ${v_w} = 100\,\rm km\,{s^{ - 1}}$ and the wind is confined but could extend up to $R_w \sim 10^{15} \,\rm cm$ \citep{yaron17}. We assume the wind starts to exist from the radius of the stellar envelope $r_* $ with a typical value around hundreds of solar radius. At a radius $R$ $(R>r_*)$ in the wind, the energy of shock swept-up wind material is ${E_s} = v_s^2\int_{{r_ * }}^{R} {4\pi {r^2}\rho dr \simeq 4\pi AR v_s^2} $, where $v_s$ is the shock velocity. The energy of shocked wind is given by the SN ejecta with velocity $v>v_s$, so one has the dynamical evolution of the shock speed in the wind by making ${E_s}({v_s}) = {\left. {E( > v)} \right|_{v = {v_s}}}$ \citep{li18},

\begin{equation}
{v_s} = {\left( {\frac{{{E_k}v_b^6}}{{4\pi A}}} \right)^{1/8}}{R^{ - 1/8}} = 6.9 \times {10^8}\,R_{15}^{ - 1/8}{\mathscr{A}^{ - 1/8}}{\mathscr{E}^{1/2}}{\mathscr{M}^{ - 3/8}}\,\rm cm \,s^{-1}.
\label{vs}
\end{equation}
Note that the above equation is available for $v_s > v_b $ while if $v_s< v_b$ the dynamical evolution should be derived by making $E_s=E_k$. In the situation considered here, Eq.~\ref{vs} is always valid.

The shock precursor has a characteristic optical depth, $\tau_c=c/v_s$, estimated by equating radiation diffusive velocity and the shock velocity. The shock breakout happens when the optical depth of wind material ahead of the shock is $\tau_w=\tau_c$, where at radius $R$ one has $\tau_w\simeq(\rho/m_p)\sigma_T R$. As a result, the shock breakout radius can be written as ${R_{br}} = 2.2 \times {10^{13}}{\mathscr{A}^{7/9}}{\mathscr{E}^{4/9}}{\mathscr{M}^{ - 1/3}}\,\rm cm.$
If this radius is smaller than the size of stellar envelope, i.e., $R_{br}  \le r_*$, the shock breakout will take place on the stellar surface. Hereafter, we adopt $r_*=R_{br}$ for simplification.

\section{Particle Acceleration and Energy Loss}\label{s3}

The SN shock may be radiation-mediated when the optical depth of Thomson scattering $\tau>\tau_c$ so that the particle acceleration is prohibited \citep{katz11, murase11}. The particle acceleration could be executed once the radiation start to escape and the shock is expected to be collisionless, say, at $R>R_{br}$ \citep{waxman01, waxman17}. The differential proton density accelerated and injected at the radius $R$ ($R>R_{br}$) is assumed to be a power-law with a highest energy exponential cutoff,
\begin{equation}
N_p^{inj}({E_p},R) = {N_0}(R)E_p^{ - s}\exp ( - {E_p}/{E_{p,\max}}).
\end{equation}

In order to find the highest cutoff energy of proton $E_{p,\max}$, one needs to evaluate the acceleration timescale, the dynamical timescale and the cooling timescales of proton. The magnetic field strength in the shocked wind can be estimated by $B = \sqrt {8\pi {\epsilon _B}\rho v_s^2} =13.5 \epsilon _{B, - 2}^{1/2}R_{15}^{ - 9/8}{\mathscr{A}^{3/8}}{\mathscr{E}^{1/2}}{\mathscr{M}^{ - 3/8}}\,\rm G$, where $\epsilon _B$ is the equipartition parameter of the magnetic energy with a typical value $\epsilon _B=0.01 \epsilon _{B,-2}$. The shock acceleration timescale is given by ${t_{acc}} = \kappa {E_p}/\beta _s^2 eBc$, where $\beta_s =v_s/c$, and $\kappa$ indicates the uncertainty of the acceleration theory. To explore the maximum energy of proton broadly, we adopt two values, i.e., $\kappa=20/3$ and $\kappa=1$ for the Bohm diffusion and some theoretical prediction beyond the Bohm limit (e.g., \cite{malkov06}). The dynamical timescale is $t_{dyn} \simeq R/v_s$. The timescale of $pp$ cooling can be given by $t_{pp}={\left[ {{\rm{0}}{\rm{.5}}{\sigma _{pp}}n_{sw} c} \right]^{{\rm{ - 1}}}}$, where $\sigma _{pp}$ is the cross section of $pp$ collision, $n_{sw}=\rho_{sw}/m_p$\footnote{We neglect the contribution of possible He abundance.} is the number density of the shocked wind and $\rho_{sw}=4 \rho$. The timescale of proton synchrotron cooling in the magnetic field $B$ is $t_{syn}=9({\gamma _p} - 1)m_p^3{c^5}/4{e^4}{B^2}\gamma _p^2\beta _p^2$, where $\gamma _p=E_p/m_p c^2$ is the Lorentz factor of proton and $\beta_p$ is the velocity of proton in unit of light speed.

\begin{figure}
\epsscale{.96} \plotone{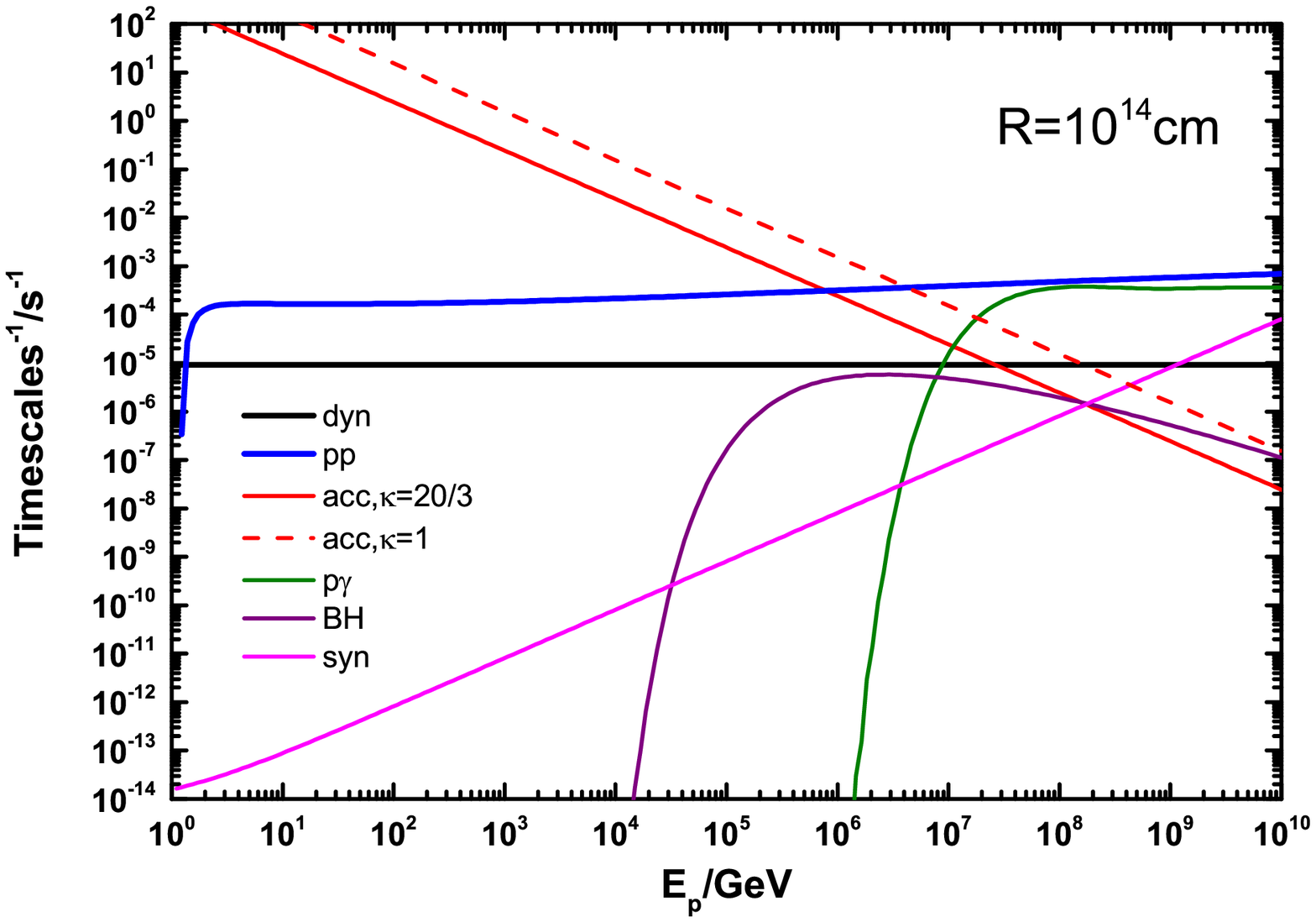} \caption{The timescales of proton in the shocked wind at radius $R=10^{14}\,\rm cm$}
\label{f1}
\end{figure}

\begin{figure}
\epsscale{.96} \plotone{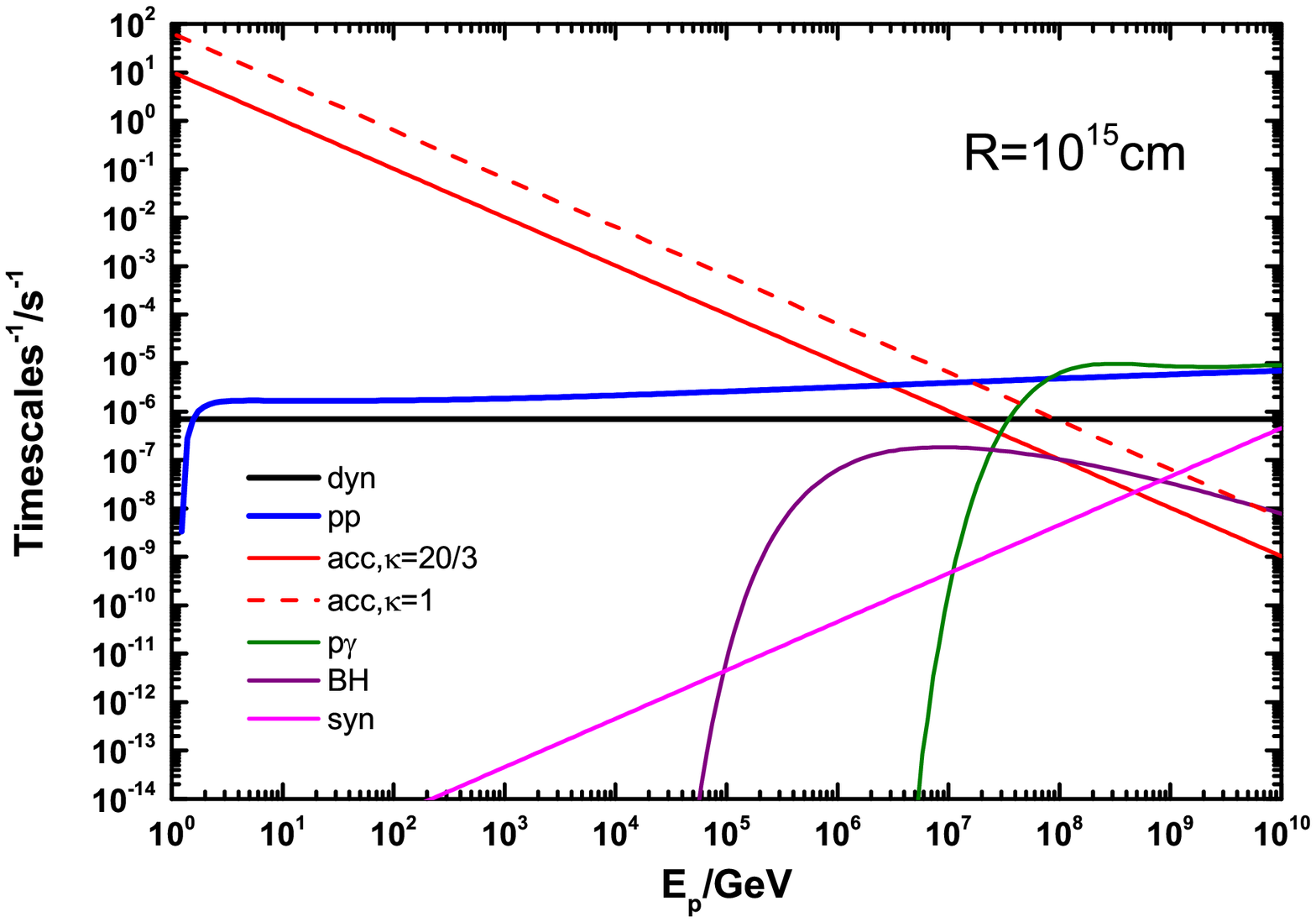} \caption{The timescales of proton in the shocked wind at radius $R=10^{15}\,\rm cm$}
\label{f2}
\end{figure}

In addition, the other cooling processes of proton related to low-energy radiation field, e.g., the photomeson ($p\gamma$) interaction and the Bethe-Heitler process (BH, $p + \gamma  \to p + e + {e^ + }$) are taken into account. For the regular Type II SN 2013fs-like case, the estimated bolometric luminosity based on the multiband photometry is around few$\times 10^{42}\,\rm erg/s$ and the blackbody temperature is around few$\times 10^{4}\,\rm K$ \citep{yaron17}. In this work, the low-energy photon filed is adopted as a blackbody distribution with a temperature $kT=2 R_{14}^{-1/2}\,\rm eV$ so that the bolometric luminosity can be around the observational value. The photospheric radius can be evaluated by $\tau(R_{ph})=1$, where $\tau (R) = \int_R^{{R_w}} {{\sigma _T}(\rho /{m_p})dr} $ is the optical depth of the materials from $R$ to $R_w$, so one can obtain $R_{ph}\simeq 6 \times 10^{14}\,\mathrm{cm} \sim R_w$, which indicates the assumption of blackbody distribution of low-energy photons is approximately valid in the considered situation. All relevant timescales are plotted in Fig.~\ref{f1} and Fig.~\ref{f2} for two representative radius $R=10^{14}\,\rm cm$ and $R=10^{15}\,\rm cm$, respectively. $E_{p,\max}$ can be obtained by letting ${t_{acc}}=\min (t_{dyn},t_{pp},t_{syn},t_{p\gamma},t_{BH})$. As we can see in two figures, $E_{p,\max}$ is mainly determined by the timescale of $pp$ collision and below $E_{p,\max}$ the main energy loss process is always $pp$ cooling in the adopted parameters. The photomeson and BH processes tend to be neglected as they are basically operated at higher energy than $E_{p,\max}$. Owing to ${t_{pp}} \propto \rho _{sw}^{ - 1} \propto {R^2}$ and ${t_{dyn}} \propto {R^{9/8}}$, at the smaller radius the $pp$ cooling would be more dominant than the dynamical evolution.

The time-dependent (or radius-dependent) energy injection rate of the shocked wind can be given by $L_{sw}=4\pi R^2 u_{sw}v_s$, where $u_{sw}=\frac{1}{{4}}\rho_{sw} v_s^2$ is the energy density of the swept-up wind by the shock. The accelerated protons typically carry a fraction $\xi=0.1\xi_{-1}$ of the shock energy, i.e., $L_p=\xi L_{sw}$, so one has $L_p=4\pi \xi A v_s ^3$. As a result, the energy density of accelerated protons can be described by $u_p^{inj}(R)=L_p/4 \pi R^2 v_s =\xi v_s^2 \rho(R)$. The normalization factor $N_0(R)$ of the distribution of the injected proton can be derived by
\begin{equation}
u_p^{inj}(R) = \int {{E_p}N_p^{inj}({E_p},R)d{E_p}}.
\end{equation}

\section{Gamma-ray and Neutrino Production}\label{s4}

In our numerical calculations, the distribution of secondaries of $pp$ collisions is obtained by following the semi-analytical method provided by \cite{kelner06}. The detailed treatment of secondaries from $pp$ collision could be found in the Appendix. Denote the emissivity of gamma-rays or neutrinos as ${N_i}({E_i}) = {\mathscr{F}_i}\left\{ {{N_p}({E_p}),{n_{sw}}} \right\}$ by invoking a operator $\mathscr{F}_i$, where $i=\gamma$ or $\nu$. Suggested by \cite{liu18a}, if we consider a group of protons with a distribution of $N_p ^{inj}(E_p,r)$ injected at a radius $r$, when they propagate to a radius $R$ the differential number density is changed to
\begin{equation}
{N_p}({E_p},r;R) = N_p^{inj}({E_p},r)\exp[ - (1 - {2^{1 - s}}){\tau _{pp}}({E_p},r,R) - (s - 1){\tau _{ad}}(r,R)].
\end{equation}
Here, the main energy loss processes as shown in Fig.~\ref{f1} and Fig~\ref{f2}, i.e., the $pp$ collision and the adiabatic cooling, are taken into account during the propagations of protons. ${\tau _{pp}}({E_p},r,R) = {\sigma _{pp}}({E_p})c\int_{t(r)}^{t(R)} {{n_{sw}}(\tilde r)dt} $ indicates the optical depth of $pp$ collision of protons injected at $r$ propagating to $R$. ${\tau _{ad}}(r,R) = \int_r^R {{v_s}(\tilde r)} dt/\tilde r = \int_r^R {d\tilde r/\tilde r}  = \ln (R/r)$ is related to the adiabatic cooling of proton moving from $r$ to $R$.  The differential luminosity of secondaries through the $pp$ collision for the shock front at $R$ can be derived by integrating over all radius ($r_*<r<R$), i.e.,
\begin{equation}
{L_i}({E_i},R) = E_i^2\int_{{r_ * }}^R {{\mathscr{F}_i}\left\{ {{N_p}({E_p},r;R),{n_{sw}(R)}} \right\}4\pi {r^2}dr}.
\label{lunu}
\end{equation}

The high-energy gamma-rays produced by $pp$ interactions would be attenuated by the low-energy photon field through $\gamma+\gamma \to e +e^+$ and absorbed by the low-energy proton through the BH process in the emission region \citep{murase11}. The gamma-rays escaped from the emission region should be multiplied by a factor of $\left[ {1 - \exp ( - {\tau _{\gamma \gamma }} - {\tau _{\rm BH}})} \right]/({\tau _{\gamma \gamma }} + {\tau _{BH}})$, where ${\tau _{\gamma \gamma }}({E_\gamma }) \simeq R\int {{\sigma _{\gamma \gamma }}({E_\gamma },\varepsilon ){N_\varepsilon }(\varepsilon )d\varepsilon }$ and ${\tau _{\rm BH}} \simeq R{\sigma _{\rm BH}}{n_{sw}}$. Two optical depths are calculated numerically in this work and for simplicity the cross section of BH process, ${\sigma _{\rm BH}}$, is adopted approximately as a fixed value $10\,\rm mb$. The low-energy photon field, ${N_\varepsilon }$, is assumed as a blackbody distribution as adopted above. Besides, the very high-energy photons will be attenuated due to the cosmic microwave background (CMB) and extraglactic background light (EBL) by a factor $e^{-\tau_{\rm CMB}-\tau_{\rm EBL}}$. The model of EBL is based on \cite{finke10}. Note that we neglect the contribution of secondary electrons produced by $pp$ collisions even though the highest energy electrons may radiate $\sim \rm GeV$ photons by synchrotron radiation in the adopted magnetic field. This is because the secondary electrons that can contribute $\sim \rm GeV$ photons are produced by the protons with energies around the cutoff energy $E_{p,\max}$, where the luminosity of protons is already significantly smaller than that of relatively low energy protons for a index $s=2$ or softer due to a exponential cutoff. Another reason is that the emissivity of gamma-rays is about two times of that of electrons during the $pp$ interaction so that the synchrotron of electron at GeV band is subdominant.

\begin{figure}
\epsscale{.96} \plotone{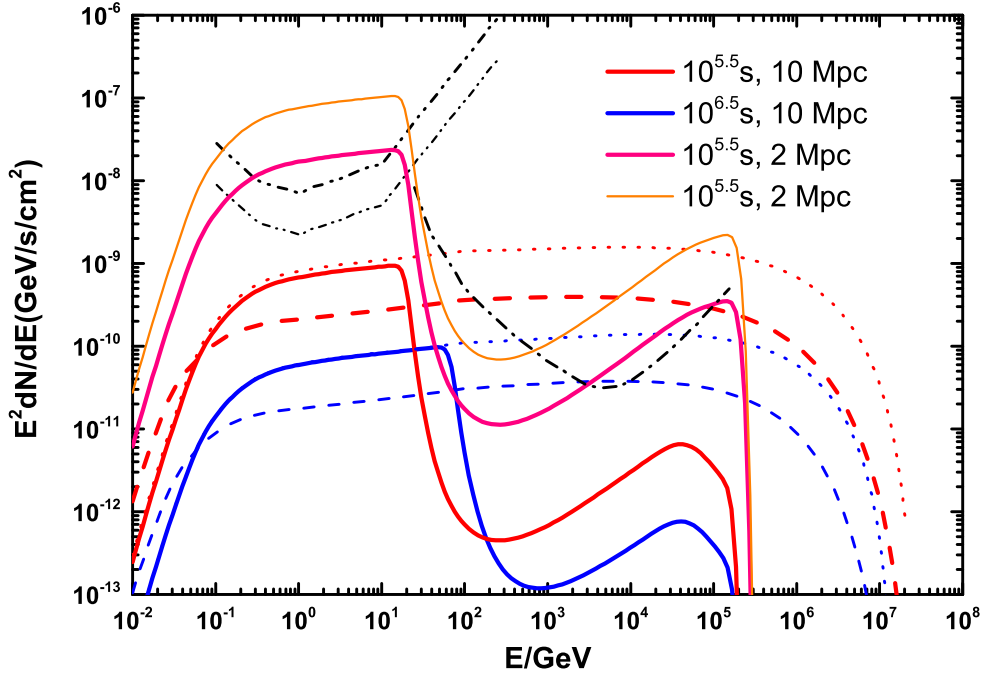} \caption{The fluxes of produced gamma-rays (solid lines) and neutrinos (dashed lines). For reference, the dotted lines indicate the gamma-ray flux without absorption. The thick and thin dot-dot-dashed lines represent respectively the differential sensitivity of \emph{Fermi}/LAT for a observational time $10^{5.5}\,\rm s$ and $10^{6.5}\,\rm s$, the dot-dashed line indicates the $50\,\rm hr$ differential sensitivity of CTA. Here, the parameters with typical values are involved, i.e., $\epsilon_B=0.01$, $\xi=0.1$, $\mathscr{A}=\mathscr{E}=\mathscr{M}=1$ and $\kappa=20/3$ except for the orange thin solid line we show the result for a denser wind environment ($\mathscr{A}=3$).}
\label{f3}
\end{figure}

The gamma-ray and neutrino production are presented in Fig.~\ref{f3}. As we can see, the gamma-ray emissions above $\sim 10 \,\rm GeV$ would be suppressed significantly by the low-energy blackbody photon field, so a different setup of low-energy photon field could make a different gamma-ray flux. In this work, the low-energy photon filed for a regular Type II SN is based on the observations of SN 2013fs. Due to the absorption of low-energy photon field in the shocked wind, the spectrum present a significant suppression at the energy range $\sim 10\,\mathrm{GeV}-100\,\mathrm{TeV}$, while the influence of the absorption of BH process is very weak, which can be slightly seen (the difference of the red solid line and the red dotted line below $10\,\rm GeV$ in Fig.~\ref{f3}) at the early stage when the density of low-energy proton is high. In Fig.~\ref{f3}, the sensitivities of \emph{Fermi}/LAT and CTA (Cherenkov Telescopes Array) are shown to compare with the gamma-ray emissions. At the radius $R=10^{14}\,\rm cm$, the duration of emission is $t_d \simeq R/v_s \sim 10^{5.5}\,\rm s$, while at radius $R=10^{15}\,\rm cm$ one has $t_d \sim 10^{6.5}\,\rm s$. For typical values of parameters, i.e., $\epsilon_B=0.01$, $\xi=0.1$, $\mathscr{A}=\mathscr{E}=\mathscr{M}=1$, at $10\,\rm Mpc$, the high-energy gamma-rays is hard to be observed by the current and next-generation telescopes for a 2013fs-like case. However, at a distance $ \lesssim 2-3 \,\rm Mpc$ the gamma-rays around GeV could be detected by \emph{Fermi}/LAT and the gamma-rays around few$-100 \,\rm TeV$ could be detected by the CTA. Note that either through the early-time spectra modeling (e.g., in \cite{yaron17}) or through the early-time lightcurves modeling (e.g., in \cite{forster18}), basically, one can only obtain the density, the profile and the extended radius of wind, while the mass-loss rate $\dot M $ and the mass-loss duration $t_w$ before the SN explosion are estimated by assuming a wind velocity $v_w$ \citep{morozova17}. In \cite{yaron17}, they achieve $\dot M =3\times 10^3\,\rm {M_ \odot }\,{yr^{ - 1}}$ by assuming $v_w =100\,\rm km/s$, while in \cite{forster18}, $\dot M $ is with a comparable value but $v_w $ is much smaller (the terminal wind velocity is assumed as $10 \,\rm km/s$), indicating a much larger density of wind (i.e., a larger $\mathscr{A}$). To explore the gamma-ray radiation broadly, we also tried a larger $\mathscr{A}$. For a denser wind environment (e.g., $\mathscr{A}=3$ shown by the orange thin solid line in Fig.~\ref{f3}), the flux of gamma-rays is significantly enhanced and it could be still detectable for a further distance of source.

\begin{figure}
\epsscale{.96} \plotone{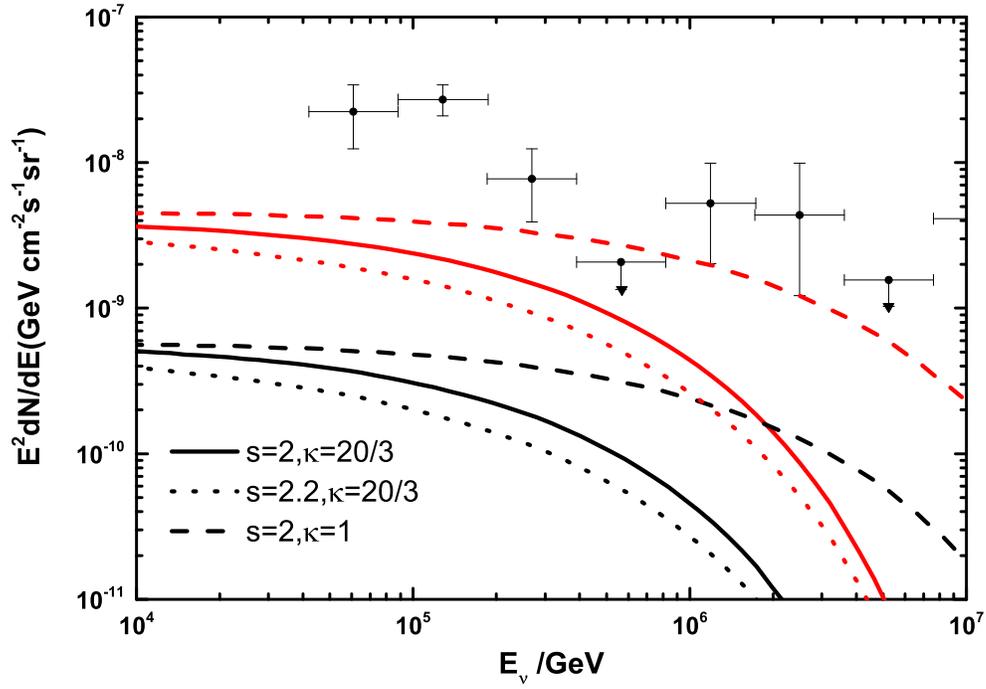} \caption{Diffuse neutrino flux (per flavor). The black solid, dotted and dashed lines represent the contribution from wind breakouts of SNe II under the same parameters as in Fig.~\ref{f3} except for the parameters shown in the figure. The red lines indicate the corresponding flux for a denser wind environment $\mathscr{A}=3$. The data of diffuse neutrinos flux is taken from \cite{aartsen17a}.}
\label{f4}
\end{figure}

The diffuse neutrino intensity from all SNe II wind breakouts in the universe can be given by integrating the contributions of individual wind breakout event at different cosmological epochs,
\begin{equation}
\frac{{d\phi }}{{dE}} = c\int {R(z)\frac{{dN}}{{dE'}}} (1 + z)\frac{{dt}}{{dz}}dz,
\end{equation}
where $dN/dE' = \int_{{t_ * }(r_*)}^{{t_w}(R_w)} {{L_\nu }(E',t)/{{E'}^2}dt} $ with $E' = E(1 + z)$. $dN/dE'$, this term dedicate to express the total neutrino production for a individual wind breakout event and ${L_\nu }(E',t)$ could be found by Eq.~\ref{lunu}. Also, $dz/dt = {H_0}(1 + z){\left[ {{\Omega _M}{{(1 + z)}^3} + {\Omega _\Lambda }} \right]^{1/2}}$ and we adopt ${\Omega _M} = 0.27$, ${\Omega _\Lambda } = 0.73$ and ${H_0} = 70\,\rm km/s/Mpc$ in our calculations. $R(z) = R(0)S(z)$ is the SNe II event rate at redshift $z$, where $R(0)$ is the local event rate and $S(z)$ is the redshift evolution of event rate that is assumed to follow the star formation rate \citep{yuksel08}. The volumetric rates of nearby core-collapse SNe is measured as $0.7 \times {10^{ - 4}}\,\rm Mp{c^{ - 3}}y{r^{ - 1}}$ \citep{li11}, most of which are SNe II. Our result is presented in Fig.~\ref{f4}. By assuming that all SNe events are 2013fs-like, the diffuse neutrino flux from wind breakouts of SNe II is subdominant in the diffuse neutrino detected by IceCube with a contribution around few percent. However, if the wind environment is denser, e.g., $\mathscr{A}=3$, and the maximum energy of accelerated proton is optimistic, i.e., $\kappa=1$, the contribution of the wind breakouts of SNe II to diffuse neutrino could be conspicuous above $300\,\rm TeV$. The diffuse neutrino flux obtained in the numerical calculations is consistent with the analytical estimation of \cite{li18}. The detailed contribution to the diffuse neutrino flux is up to the spectral index of protons as well and a softer distribution of protons will make the contribution slightly less.

\section{Discussion and Conclusion}\label{s5}

\subsection{High-energy gamma-rays and neutrinos}
In this work, we have studied the gamma-ray and neutrino emission during the interaction of SN ejecta with the dense wind, which may come out almost simultaneous with the optical/infrared lights. For a SN 2013fs-like wind, the ratio of shock velocity to the bulk velocity is $v_s/v_b \simeq 2.2 R_{15}^{ - 1/8}{\mathscr{A}^{ - 1/8}}{\mathscr{M}^{1/8}}$, so one obtains the fraction of shock energy in the bulk ejecta energy,
\begin{equation}
\eta  = {\left. {E( > {v_s})} \right|_{v = {v_s}}}/{E_k} = {({v_s}/{v_b})^{ - 6}} = 9 \times {10^{ - 3}}R_{15}^{3/4}{\mathscr{A}^{3/4}}{\mathscr{M}^{ - 3/4}}.
\end{equation}
As a result, the wind breakouts of SN II shocks can convert a fraction $\eta \xi \simeq 9 \times 10^{-4} \xi_{-1} R_{15}^{3/4} {\mathscr{A}^{3/4}}{\mathscr{M}^{ - 3/4}}$ of the bulk energy into accelerated protons. The accelerated protons undergo the significant cooling by $pp$ interactions and transfer almost total energy to secondaries. For gamma-rays, under the typical parameters of 2013fs-like case, the $\sim\,\rm GeV$ and $\sim \mathrm{few}-100 \,\rm TeV$ gamma-rays could be detected at $\lesssim 2-3\,\rm Mpc$ by \emph{Fermi}/LAT and CTA during the ejecta-wind interaction, respectively. For the SN II wind breakout as the point source of neutrino, at $10\,\rm Mpc$, the flux is $\sim 3\times 10^{-10}\,\rm GeV cm^{-2} s^{-1}$, which could reach the sensitivity level of future IceCube Gen2 \citep{aartsen17b}, and at closer distance or a galactic event, the neutrinos could be detected by current IceCube \citep{murase18a}. Furthermore, the efficiency of $pp$ interaction is proportional to the number density of wind, i.e., $\propto  \mathscr{A}$, so the fluxes of secondaries is proportional to $\mathscr{A}^{7/4}$. Consequently, if a SN is with a denser wind environment ($\mathscr{A}>1$), it could be still detectable with a further distance.


By assuming all SNe II are 2013fs-like, we have presented the per-flavor diffuse neutrino flux from SN II wind breakouts is $\sim 5 \times 10^{-10}\,\rm GeV cm^{-2} s^{-1} sr^{-1}$ or with a contribution about few percent of the observed diffuse neutrino flux, which is smaller than estimated diffuse neutrino flux from SNe IIn even though the event rate of regular SN II is much larger than that of SN IIn \citep{petropoulou17}. One possible reason is that here we consider a SN ejecta with a steep velocity distribution so that the fraction of total ejecta energy converting to shock is quite small. However, for a denser wind, e.g., $\mathscr{A}=3$, the diffuse neutrino flux from wind breakouts could reach a comparable level with the observed IceCube diffuse neutrinos above $300\,\rm TeV$. Moreover, under the assumption that the low-energy photon field in optical/infrared energy band is with a luminosity few$\times 10^{42}\,\rm erg/s$, the emitted gamma-rays with energies from tens of GeV to tens of TeV are mainly significantly absorbed in the emission region. So in this case, the accompanying diffuse gamma-ray emission with diffuse neutrino emission can be estimated as $\sim 1 \times 10^{-9}\,\rm GeV cm^{-2} s^{-1} sr^{-1}$ without considering the cascade in the intergalactic space. Such a diffuse gamma-ray flux is typically lower than that of the diffuse isotropic gamma-ray background \citep{ackermann15}.

For the high-energy gamma-rays, the \emph{Fermi}/LAT and CTA are able to detect the signatures of the wind breakouts of Type II SNe at $2-3\,\rm Mpc$ for a time window of several days. Such a size is comparable with the size of local galaxy cluster. The expected SN II event rate in local galaxy cluster is $\sim$few in ten years \citep{mannucci08}. The searching of accompanying gamma-rays for past nearby Type II SNe located in the FoV (Field of view) of \emph{Fermi}/LAT could be a test of wind breakout, and a follow-up observation by Fermi/LAT and CTA in the future is encouraging.

\subsection{Lower energy radiations}
In addition to the high-energy gamma-rays, next, we want to give a brief discussion about the radiations in other wavelengths. For $\tau_w \lesssim \tau_c \equiv c/v_s$, the shock is expected to be collisionless, and the energy of the shock is $\eta E_k \sim 10^{49}\,\rm erg$, only $\xi=0.1 \xi_{-1}$ of which is assumed to be converted to the relativistic particles. Thus, most of the energy of the shock is the thermal energy. The temperature of the thermal proton at the immediate downstream of shock can be estimated as $kT_p=3 m_p v_s^2/16 \simeq 93\,\rm keV$ for the typical values of relevant items in this work. The electron temperature should be not larger than the equipartition temperature $(\simeq 47\,\rm keV)$ but it is still uncertain due to the unknown efficiency by which protons transfer energy to electrons in collisionless shocks. However, since the collisionless shock heating is typically faster than Coulomb collisional processes \citep{katz11}, a lower limit for the electron temperature can be obtained by assuming the shock is collisional (in other words, there is no collisionless heating). In the absence of collisionless shock heating, the electron temperature is achieved by the balance between Coulomb heating and cooling processes. If the fastest cooling process is the inverse Compton scattering off the radiation field, one has $k{T_e} \sim 40  U_{\gamma,3} ^{ - 2/5} n_{sw,10}^{2/5}(kT_p/100 \, \mathrm{keV})^{2/5}\,\rm keV$ \citep{waxman01, katz11, murase11}, where ${U _\gamma } = L/(4\pi R^2 v_{d}) \simeq 10^3 L_{43} R_{15} ^{-2} v_{d,9}^{-1}\,\rm erg \, cm^{-3}$ is the energy density of the low-energy radiation field and $v_{d} =c/\tau_w$ is the diffusion velocity of light. Consequently, the X-rays can be naturally expected for the electrons with energies of tens of keV via inverse Compton or thermal bremsstrahlung \citep{chevalier12, Pan13}. Since the energy of electron is from Coulomb heating of proton, the radiation efficiency of shocked materials can be obtained by comparing the proton cooling timescale $t_p \sim 2 \times 10^4 U_{\gamma,3} ^{3/5}n_{sw,10}^{-8/5} (kT_p/100 \, \mathrm{keV})^{-3/5}  \,\rm s$ with the dynamical timescale $R/v_s \sim 10^6 \,\rm s$ \citep{katz11}. As a result, the cooling of the shocked materials could be efficient and contribute the thermal X-rays, the luminosity of which can be about $(1-\xi)/\xi$ of that of non-thermal gamma-rays if we neglect the external absorption of them, i.e., $\sim 10^{43}\,\rm erg/s$.

Besides, the relativistic electrons including secondary electrons (from $pp$ collisions) and primary electrons (co-accelerated with protons by the shock) can contribute to the non-thermal X-ray, radio and MeV gamma-ray emissions, and the electromagnetic cascade initiated by the absorbed high-energy gamma-rays in the emission region can give a contribution as well. The accurate calculations of them are challenged since the inelastic Compton scattering by the thermal electrons, as well as other complexities proposed in \cite{waxman17}, plays a crucial role in determining the distributions of electrons and photons but it is not in general in thermal equilibrium. However, at X-ray energy band, the non-thermal contributions tend to be subdominant because the radiations of thermal electrons are efficient and the energy of thermal electrons are typically larger than that of relativistic electrons as we mentioned above that most of energy is still thermal energy.
The radio emission could arise from secondary and primary electrons, but it may be suppressed by free-free absorption, synchrotron self-absorption and Razin-Tsytovich process, and modified by Comptonization of thermal electrons \citep{murase14}. The soft X-rays are expected to be up-Comptonized and the gamma-rays to be degraded by thermal electrons to some extent, depending on the opacity of Compton scattering. The typical photon energy may be comparable to the thermal electrons, i.e., a few tens of keV.  The soft X-ray and radio emissions have been reported in some SNe II (e.g., \cite{pooley02, chevalier06}, and references therein), but their luminosities are usually weak, ranging from $10^{37}$ to almost $10^{42} \,\rm erg/s$ \citep{dwarkadas14}, maybe implying suppression due to Comptonization.
Although in this work we mainly focus on the high-energy gamma-ray emission and the detailed discussions of X-ray and radio emissions are beyond the scope of this work, in the future, in addition to the gamma-rays, the observational constraints on X-ray and radio emissions could be helpful to check the ejecta-wind interaction model for the regular SNe II and provide the property of wind environment. In particular, X-ray missions such as HXMT \citep{xie15} and Einstein Probe \citep{yuan15} may significantly improve the prospects for the detection of accompanying X-ray emission, which, in addition to the high-energy radiations, will help us to understand the progenitor nature of SNe II.

\acknowledgments
This work is supported by the NSFC grant 11773003 and the 973 program grant 2014CB845800.

\clearpage

\appendix

\section{The secondaries produced by $pp$ collisions}

Basically, we follow the semi-analytical method provided by \cite{kelner06} (see also \cite{erwin14, liu18b}). The differential production in unit energy and unit time is given by
\begin{equation}
{\mathscr{F}_i}({E_i}) = c{n_{sw}}\int_{{E_i}}^\infty  {{\sigma _{pp}}({E_p}){N_p}({E_p})} {F_i}\left( {\frac{{{E_i}}}{{{E_p}}},{E_p}} \right)\frac{{d{E_p}}}{{{E_p}}},
\end{equation}
where $i$ could be $\gamma$ or $\nu$, and the cross section ${\sigma _{pp}}({E_p}) = 34.3 + 1.88L + 0.25{L^2}\,\rm mb$ with $L=\ln(E_p/1\,\rm TeV)$. $F_i$ is the spectrum of secondary $\gamma$ or $\nu$ in one collision, which can be found in Eqs.~58, 62, 66 of \cite{kelner06}. The above analytical presentation works for $E_p > 100 \,\rm GeV$, while for $E_p < 100 \,\rm GeV$ the spectra of secondaries can be continued to low energies using the $\delta-$functional approximation for the energy of produced pions \citep{aharonian00}, say,
\begin{equation}
{\mathscr{F}_i}({E_i}) = 2c{n_{sw}}\frac{{\tilde n}}{{{K_\pi }}}\int_{{E_{i,\min }}}^\infty  {{\sigma _{pp}}({m_p} + \frac{{{E_\pi }}}{{{K_\pi }}}){N_p}({m_p} + \frac{{{E_\pi }}}{{{K_\pi }}})} \frac{{d{E_\pi }}}{{\sqrt {E_\pi ^2 - m_\pi ^2} }},
\end{equation}
where $E_{\pi}$ is the energy of pions and the rest mass of pion $m_\pi \simeq 135\,\rm MeV$ for gamma-ray production and $m_\pi \simeq 140\,\rm MeV$ for neutrino production. ${E_{i,\min }} = {E_i}/{\varsigma _i} + {\varsigma _i}m_\pi ^2/4{E_i}$ with ${\varsigma _\gamma }=1$ and ${\varsigma _\nu } = 1 - m_\mu ^2/m_\pi ^2 = 0.427$, $K_\pi=0.17$, and ${\tilde n}$ is a free parameter that is determined by the continuity of the flux of the secondaries at $100\,\rm GeV$. At lower energies one can use a more accurate approximation for the inelastic cross section of $pp$ interaction instead, i.e., ${\sigma _{pp}}({E_p}) = (34.3 + 1.88L + 0.25{L^2}){\left[ {1 - {{\left( {{E_{th}}/{E_p}} \right)}^4}} \right]^2}\,\rm mb$ with $E_{th}=1.22\,\rm GeV$.

\clearpage


\end{document}